\documentclass[%
 aip,
 amsmath,amssymb,
 reprint,%
]{revtex4-2}

\usepackage[utf8]{inputenc}
\usepackage[T1]{fontenc}
\usepackage{mathptmx}
\usepackage{etoolbox}

\makeatletter
\def\@email#1#2{%
 \endgroup
 \patchcmd{\titleblock@produce}
  {\frontmatter@RRAPformat}
  {\frontmatter@RRAPformat{\produce@RRAP{*#1\href{mailto:#2}{#2}}}\frontmatter@RRAPformat}
  {}{}
}%
\makeatother

\usepackage{graphicx}
\graphicspath{{./figs/}}
\usepackage{dcolumn}
\usepackage{bm}
\usepackage{amsmath}
\usepackage{xcolor}
\usepackage[linesnumbered,ruled]{algorithm2e}
\usepackage[braket, qm]{qcircuit}
\usepackage{tikz}
\usetikzlibrary{shapes.geometric, arrows}
\usepackage{svg}

\usepackage{xr}
\makeatletter
\newcommand*{\addFileDependency}[1]{
\typeout{(#1)}
\@addtofilelist{#1}
\IfFileExists{#1}{}{\typeout{No file #1.}}
}\makeatother

\newcommand*{\myexternaldocument}[1]{%
\externaldocument[#1-]{#1}%
\addFileDependency{#1.tex}%
\addFileDependency{#1.aux}%
}
\myexternaldocument{Supp}
\input{Supp.aux2}

\begin{document}

\title[Hamiltonian Partitioning]{Optimised Baranyai partitioning of the second quantised Hamiltonian}

\author{Bence Csakany*}
    \email{bc528@cam.ac.uk}
\author{Alex J.W. Thom}%
\affiliation{ 
Yusuf Hamied Department of Chemistry, Cambridge, U.K.
}%

\date{\today}

\begin{abstract}
Simultaneous measurement of multiple Pauli strings (tensor products of Pauli matrices) is the basis for efficient measurement of observables on quantum computers by partitioning the observable into commuting sets of Pauli strings. We present the implementation and optimisation of the Baranyai grouping method for second quantised Hamiltonian partitioning in molecules up to CH\textsubscript{4} (cc-pVDZ, 68 qubits) and efficient construction of the diagonalisation circuit in $O(N)$ quantum gates, compared to $O(N^2)$\cite{gokhale2019GeneralCommute}, where $N$ is the number of qubits. We show that this method naturally handles sparsity in the Hamiltonian and produces a $O(1)$ number of groups for linearly scaling Hamiltonians, such as those formed by molecules in a line; rising to $O(N^3)$ for fully connected two-body Hamiltonians. While this is more measurements than some other schemes\cite{inoue_almost_2023} it allows for the flexibility to move Pauli strings between sets and optimise the variance of the expectation value\cite{crawford_efficient_2021}.  We also present an explicit optimisation for spin-symmetry which reduces the number of groups by a factor of $8$, without extra computational effort. 
\end{abstract}

\maketitle


\section{Introduction}
Complete measurement on a quantum computer yields a bitstring that corresponds to the measured state of each individual qubit. Repeated measurement yields bitstrings with frequency proportional to the squared amplitude of the statevector component corresponding to that bitstring. For $N$ qubits there are $2^N$ possible bitstrings and the statevector has dimension $2^N$.
Measurement of the expectation value of an operator acting on a Hilbert space corresponds to the matrix product: 
\begin{equation}
\langle \hat{O} \rangle = \langle \Psi | O | \Psi \rangle
\label{eq:Measurement}
\end{equation}
where $\Psi$ represents the statevector, in which each bitstring addresses an element of the Hilbert space and $O$ corresponds to the matrix representation of the operator $\hat{O}$. 

To measure efficiently requires expressing the operator matrix as the sum of implementable and diagonalisable matrices. Generally Pauli strings ($N$ dimensional tensor products of Pauli matrices) are chosen as they form a basis for the complex vector space of $2N\times 2N$ matrices. These Pauli strings are then diagonalised and, using an estimate of the statevector amplitudes, the expectation value can be computed. 

This procedure is detailed in Section \ref{sec:OperatorExpression} and for matrices which can be simultaneously diagonalised only one estimate of the statevector amplitudes is required, therefore grouping matrices into simultaneously diagonalisable sets can yield large speedups.

An alternative to simultaneous measurement is the partitioning of the operator into anti-commuting sets. These can then be rotated back to only one Pauli string which can then be measured, leading to the evaluation of multiple Pauli strings\cite{zhao_measurement_2020,ralli_implementation_2021}. Shadow tomography is another alternative, where $M$ different expectation values can be estimated with $\text{log}(M)$ measurements; however, the error and efficiency depends on the locality of the Pauli strings\cite{huang_predicting_2020}. Other techniques\cite{izmaylov_revising_2019} convert the Hamiltonian into a sum of mean field Hamiltonians which can be efficiently measured. \\

There are several schemes where Pauli strings are grouped into commuting cliques are suggested in the literature\cite{bonet-monroig_nearly_2020, inoue_almost_2023, gokhaleDisjointCommute}, however in general the problem is equivalent to a min-clique cover problem which has been shown to be NP-Hard\cite{minCliqueNpHard}. Therefore, even for relatively small systems this procedure is usually too time consuming to complete optimally. Nevertheless schemes have been found that achieve an $O(N^2)$ measurement cost\cite{bonet-monroig_nearly_2020, inoue_almost_2023}.
However, while fewer sets of commuting Pauli strings yield more samples per string, for the same overall sample count, it is not guaranteed that this will minimise the variance, as is demonstrated in Section 2 of Ref \onlinecite{crawford_efficient_2021}. Several schemes have been suggested that aim to minimise the variance in the expectation value, not the number of commuting sets\cite{choi_improving_2022,crawford_efficient_2021,kohda_quantum_2022,choi_measurement_2023}. These techniques aim to distribute the Pauli strings among the measurement sets such that the overall variance is minimised, to this effect partitioning schemes which allow for this flexibility need to be used. We focus on one of these in this letter.\\
It has been shown that systems described in terms of second quantised excitation operators fall into a special case where the Baranyai partitioning can be computed in $O(N^5 \text{log}N)$ complexity\cite{gokhaleDisjointCommute}. We implement this partitioning method and extend it to incorporate spin symmetry, reducing the number of groups by a factor 8, provide an efficient diagonalisation routine, and extend it to other mappings. We demonstrate this procedure for the second quantised Hamiltonian of systems up to 68 qubits (CH\textsubscript{4} cc-pVDZ) encoded under JW (Jordan-Wigner)\cite{jordan_uber_1928} mapping. This is larger than any system which can be currently be evaluated (simulated or real) at the NISQ (Noisy Intermediate Scale Quantum) computing level.\\

Section \ref{sec:Background} describes the background necessary, section \ref{sec:Groupthem} describes the procedure for grouping, Section \ref{sec:GeneralDiag} shows the diagonalisation procedure, and Figure \ref{fig:BaranyaiGroupingResult} shows this in action for a series of molecules in several bases (STO-3G, 6-31G, cc-pVDZ).
\section{Background}
\label{sec:Background}

\subsection{Expectation Value of an Operator}
\label{sec:OperatorExpression}
To compute the expectation value $\langle \Psi | O | \Psi \rangle$ we rotate the statevector to a basis in which $O$ is diagonal: $O = U^\dag \Lambda U$. We then compute:
\[\langle \Psi | O | \Psi \rangle = \langle U\Psi | \Lambda | U\Psi \rangle\]
where we know an approximation to $| U\Psi \rangle$ from the samples measured. This does not require storing a $2^N$ statevector as the expectation value can be computed as a running total as the data is received. The same approximation of $| U\Psi \rangle$ can be used if a different $O'$ is also diagonalised by $U$. Therefore we search for maximally commuting sets of operators.

\subsection{JW Mapping}
\label{sec:JW}
To encode problems onto a quantum computer we must choose a mapping from our problem space to the qubit space. In the context of chemistry this means usually mapping the space of Slater determinants to qubits. The JW\cite{jordan_uber_1928} mapping is the most simple one, being:
\begin{gather*}
a^\dag_j = Z^{\otimes {j-1}} \otimes \frac{X-iY}{2}\\
a_j = Z^{\otimes {j-1}} \otimes \frac{X+iY}{2}
\end{gather*}
Where \(a_j, a^\dag_j\) represent the annihilation (creation) operator in the $j$\textsuperscript{th} spin-orbital respectively. This maps the entire Fock space to the qubit space. With this, excitation operators ($a^\dag_{j}a_i$) acting in the Fock space can be converted to Pauli strings which can be evaluated on a quantum computer. This definition can be interpreted as the $j$\textsuperscript{th} qubit being `1' representing the occupation of an electron in the $j$\textsuperscript{th} spin-orbital.

\subsection{Hamiltonian Operator}

The second quantised Hamiltonian operator for electronic structure problems can be expressed as (using an implicit summation over repeated indices):

\[\hat{H}=h_{ij} a_i^\dag a_j + h_{ijkl} a_i^\dag a_j^\dag a_k a_l\]

To evaluate \( \langle \psi_{\text{trial}} |\hat{H}|\psi_{\text{trial}} \rangle \) we use the JW mapping of the action of creation and annihilation operators in the Fock space to convert the Hamiltonian to a sum of Pauli strings:

\[H = \sum_i{h_i P_i}\]

where $P_i$ are the Pauli Strings generated from expanding out the excitation operators under JW mapping as in Section \ref{sec:JW} and $\psi_{\text{trial}}$ is restricted to a vector in the Fock space.

\[\langle \psi_{\text{trial}} |H|\psi_{\text{trial}} \rangle = \sum_i{c_i \langle \psi_{\text{trial}} |P_i|\psi_{\text{trial}} \rangle}\]


To evaluate each $\langle \psi_{\text{trial}} |P_i|\psi_{\text{trial}} \rangle$ we proceed as in Section \ref{sec:OperatorExpression}, where a Pauli string is diagonal if it is the tensor product of only $\{Z,I\}$ matrices.


\section{Grouping Pauli Strings}
\label{sec:Groupthem}
The choice of $U$ is not unique; we can choose up to $2^N$ Pauli strings which can be simultaneously diagonalised (where $N$ is number of qubits) as there are $2^N$ unique diagonal Pauli strings. Two Pauli strings can be simultaneously diagonalised iff they commute. A maximally commuting set forms a group under matrix multiplication, as the product of diagonal matrices is also diagonal, therefore it suffices to look for commutativity of generators of any subgroups. This notion is formalised in \cite{sarkar2019PauliCommuteSets}\cite{reggio2023fast}.\\

\subsection{Structure of the Excitation operators under JW mapping}
The structure of the excitation operators yields some helpful insights that can be exploited to produce maximally commuting sets in the general case where the Hamiltonian has no zero coefficients ($h_{ijkl} \neq 0, \forall{i,j,k,l}$). With some modifications this is a good approximation otherwise as well (see Figure \ref{fig:BaranyaiGroupingResult}).\\
Expanding a general single-excitation operator yields (for $j \neq i$):
\begin{gather*}
a_i^\dag a_j = \\
XZ^{\otimes |j-i|}X + YZ^{\otimes |j-i|}Y + i(XZ^{\otimes |j-i|}Y + YZ^{\otimes |j-i|}X)\\
\equiv XX + YY + i(XY + YX)
\end{gather*}
where using the shorthand notation, in the last line the $\{Z,I\}$ matrices are implied. See \cite{tazi2023folded} for a fuller explanation. 

\subsection{Cancellation within $\hat{H}$ }
For a real Hamiltonian, the coefficients $h_{ij}$ and $h_{ijkl}$ are real, and the Hamiltonian is hermitian, so:
\[h_{ij} = h_{ji}\ \text{and}\ h_{ijkl} = h_{klij},\]
and so the Hamiltonian can be factorised as:
\[\hat{H}=\frac{1}{2} h_{ij}(a_i^\dag a_j + a_j^\dag a_i) + \frac{1}{2} h_{ijkl} (a_i^\dag a_j^\dag a_k a_l + a_k^\dag a_l^\dag a_i a_j)\]
Therefore we only need to concern ourselves with the `Real' component of the excitation operators. 
It can be seen that the `Real' and `Imaginary' part of a single excitation operator form commuting sets of Pauli Strings:  $\{XX,YY\}$ and $\{XY,YX\}$\cite{gokhaleDisjointCommute}. 
It can be shown that the Pauli strings within excitation operators with disjoint indices commute\cite{gokhaleDisjointCommute}. See also section \ref{Supp-app:disjointCommute} in the supplementary information. This motivates the following grouping.

\subsection{Baranyai Grouping}
\label{sec:Baranyai}
Baranyai's Theorem states:\\ 
\textit{If $k$ divides $n$ then the $k$-element subsets of an $n$-element set can be partitioned into $\binom{n-1}{k-1}$ classes so that every class contains $n/k$ disjoint $k$-element sets and every $k$-element set appears in exactly one class}\cite{BARANYAI1979276}\\
In our case we choose $k$ indices from $n$ spin-orbitals, defining our excitation operator. These are then grouped as above yielding $\binom{n-1}{k-1}$ groups. This is a reduction from $O(N^4)$ double excitations in the Hamiltonian to $O(N^3)$ groups.\\

These Baranyai groups can be found by constructing a flow network and then solving with either the Ford-Fulkerson algorithm \cite{ford_fulkerson_1956} $O(N^8)$  or by flow rounding $O(N^5 \text{log}N)$, where $N$ is the number of qubits\cite{B&SFlowRoundingHypergraph,FlowRoundingStackOverflow,kang_flow_2015,gokhaleDisjointCommute}. In this work we opted for the Ford-Fulkerson method as this is more readily implementable but slower. Even so the largest systems were computed easily. To adapt this to real-world Hamiltonians it is necessary to handle the case where elements are missing, and also the case where $k$ does not divide $n$.

For the case where $k$ does not divide $n$ we can choose either to work with smaller sets, or embed it into a larger space.
Embedding into a larger space yields:
\[\binom{n+k-r-1}{k-1}\]
number of partitioned groups, where $r = n\ \text{mod}\ k$ It is shown in the supplementary information that embedding into a larger space leads to fewer partitions therefore this method was chosen. 
To handle the case of missing elements. The existing elements were ordered according to the position they would be in were all the elements present. A greedy algorithm then iterates over the elements adding them to an existing grouping if there is space, and creating new ones if there is not. This is $O(N^6)$, where $N$ is the number of qubits, complexity (which is less than the Ford-Fulkerson algorithm) and reproduces the Baranyai grouping in the limit of no missing elements. This allows the algorithm to incorporate the sparsity present in the Hamiltonian. In the limit of Linear Hamiltonians the number of groups is constant. See Figure \ref{Supp-fig:LinearHamiltonian} in Supplementary Information.

\subsection{Factorisable Double Excitations}
In the Hamiltonian, spin symmetry creates zeros in the second quantised expansion. The symmetry is such that for a given index set ${i,j,k,l}$ either 0,2,4 indices must represent spin-orbitals with $\alpha$ (or $\beta$) spin electrons. Therefore the double excitations seperate into two classes: pure $\alpha$($\beta$) doubles, These have 4(0) $\alpha$ indices respectively, and `cross' doubles. These have 2 $\alpha$ and 2 $\beta$ indices. 

$\alpha$ doubles and $\beta$ doubles always have disjoint indices, therefore we can group them separately. The $\alpha$ doubles are packed into the $\frac{N}{2}$ $\alpha$ indices, the $\beta$ doubles are packed into the other $\frac{N}{2}$ $\beta$ indices and the two groupings can be concatenated with no penalty. In the case that $\frac{N}{2}\ \text{mod}\ 4 \neq 0$, The procedure in section \ref{sec:Baranyai} is followed.\\

Cross doubles factorise by noting that for each pair of $\alpha$ indices, all pairs of $\beta$ indices can be present. This set is shown in equation \ref{eq:crossDoubles}
\begin{equation}
    \{a_i^\dag a_j^\dag a_k a_l + a_k^\dag a_l^\dag a_i a_j\ |\ k,l \in \{\beta\}\}\ \forall i,j \in \{\alpha\}
    \label{eq:crossDoubles}
\end{equation}
where $\{\alpha\}$, $\{\beta\}$ dictate the set of $\alpha$, $\beta$ indices. \\
Since once again the $\alpha$ indices are always disjoint to the $\beta$ indices we can consider them separately. We can pack $\frac{N}{4}$ cross-doubles into the $\frac{N}{2}$-sized $\alpha$ space where we have not considered what their $\beta$ indices represent. From equation \ref{eq:crossDoubles} we can see there are $\binom{N/2}{2}$ choices for the $\beta$ indices given the $\alpha$ indices, therefore this $\alpha$ space packing is repeated $\binom{N/2}{2}$ times.

To choose the $\beta$ groupings we employ Baranyai again to create packings of the $\binom{N/2}{2}$ $\beta$ indices into $\frac{N}{2}$. This generates one grouping. However, each given $\alpha$ index contains ALL possible $\beta$ indices, therefore the Baranyai packing is rotated so that these choices of two $\beta$ indices are eliminated everywhere. Repeating this for all packings of $\beta$ indices, and then all packings of $\alpha$ indices, all the cross-doubles are eliminated without repeats. An example is laid out in section \ref{Supp-sec:FactorisableDoubleExample} of the supplementary information.\\
If $N\ \text{mod}\ 8=0$, this partitions the double excitations into:
\[\binom{\frac{N}{2}-1}{3} +  \binom{\frac{N}{2}}{2} \binom{\frac{N}{2}-1}{2-1}\]
groups. giving roughly $\frac{1}{8}$ as many groups as the $\binom{N-1}{3}$ before. If $N\ \text{mod}\ 8\neq0$ embedding into a larger space yields:
\[\binom{\frac{N}{2}+4-r_4-1}{3} + \binom{\frac{N}{2}}{2} \binom{\frac{N}{2}+2-r_2-1}{1}\]
number of sets, where $r_4 = \frac{N}{2}\ \text{mod}\ 4$, $r_2 = \frac{N}{2}\ \text{mod}\ 2$. This still gives the required $\frac{1}{8}$ reduction.

\subsection{Repeated indices}
\label{sec:RepeatedIndices}
In the above we have assumed disjoint indices within an excitation operator, i.e. $i \neq j \neq k \neq l$. We will now handle the case of repeated indices.\\ Consider the case of one repeat, $j=k$. The resultant Pauli strings present, after cancellation, in a real Hamiltonian could be:
\[\{XIX,YIY,XZX,YZY\}\]
where the $Z,I$ for the repeated index has been explicitly shown in shorthand notation.
However, there are 3 choices for which index to repeat therefore the complete set of Pauli strings with a given 3 active indices is

\newcommand{\blue}[1]{\underline{#1}}
\newcommand{\red}[1]{\overline{#1}}
\begin{gather*}
    \{\blue{XIX},\red{YIY},\red{XZX},\blue{YZY}\}\\
    \{\blue{IXX},\red{IYY},\red{ZXX},\blue{ZYY}\}\\
    \{\blue{XXI},\red{YYI},\red{XXZ},\blue{YYZ}\}
\end{gather*}
These can be separated into two commuting sets, as indicated by the under(over)-lined terms, Augmenting each commuting set with $III$ and $ZZZ$, which are generated by the already present elements, shows that these sets are maximally commuting on 3 indices ($2^3 = 8$). 
Within each set the Pauli strings generated by single excitations on indices which are a subset of the 3 active indices are fully included. As a result, the partitioning of single excitations is encompassed here. \\
These can be grouped as in Section \ref{sec:Baranyai} to create $O(N^2)$ groupings and diagonalised by the procedure in section \ref{sec:GeneralDiag}
 
\section{General Diagonalisation for excitations in Baranyi grouping}
\label{sec:GeneralDiag}
There are algorithms for finding the diagonalising matrix $U$ in the general case of a commuting set\cite{gokhale2019GeneralCommute}, however due to our knowledge of the structure of the set we can find a much shorter one. 
The Clifford group is the set of matrices $V$ such that
\[VPV^\dag \in \boldsymbol{P}_n, \forall P \in \boldsymbol{P}_n\]
where $\boldsymbol{P}_n$ is the set of all n length Pauli strings\\
Therefore, we can find $U$ as an element of the Clifford group. Further, the Clifford group is generated by the quantum logic operations \{H, CNOT, Rx($\pi/2$)\} and therefore our matrix $U$ will be directly implementable. 

Table \ref{tab:CliffordGateEffect} shows the effect of conjugation by these Clifford gates on the different Pauli matrices.

\begin{table}
    \centering
    \begin{tabular}{|c|c|c|}
    \hline
        Gate & $P$ & Gate $P$ Gate$^\dag$ \\
    \hline
        H & $Z$ & $X$ \\
        Rx($\pi/2$) & $Z$ & $Y$ \\
        CNOT & $Z\otimes Z$ & $Z\otimes I$ \\
        CNOT & $I\otimes Z$ & $I\otimes Z$ \\
        CNOT & $X\otimes X$ & $I\otimes X$ \\
        CNOT & $X\otimes I$ & $X\otimes I$ \\
    \hline
    \end{tabular}
    \caption{Table showing the effect of the 3 generating Clifford operations on Pauli Strings. Y Pauli matrices are not shown as they are fully defined by the effect of X and Z. Here CNOT is defined with control on the left element of the tensor product}
    \label{tab:CliffordGateEffect}
\end{table}


The procedure to diagonalise these excitations is laid out in Algorithm \ref{alg:DiagPaulis}. The idea is to sort the Pauli strings to be diagonalised in ascending order of the largest active index where an active index is one of the indices present in the excitation generating this Pauli string. This is so that when the final $X$/$Y$ matrix is diagonalised (By a H or Rx), only non-diagonal Pauli strings are modified, thereby assuring the sequence terminates. 
An example is laid out in section \ref{Supp-sec:DiagonalisationExample} of the supplementary information\\
Each index is touched once except for index which had $Z$ matrices in other Pauli strings ($Z$ clashes). There are at most $\left(\frac{N}{n}\right)$ $Z$ clashes so the overall number of quantum circuit gates required is $O(N)$, where $N$ is the number of indices and $n$ is the number of indices per excitation. 

\begin{algorithm}
\label{alg:DiagPaulis}
    \SetKwInOut{Input}{Input}
    \SetKwInOut{Output}{Output}
    \Input{The set $P$ of Pauli strings to be diagonalised}
    \Output{The sequence of gates which diagonalises the Pauli strings}
    S = Sort($P$)\\
    U = Identity\\
    \For{Pauli string in S}
      {
        Apply U to Pauli string\\
        \If{Pauli string is not diagonalised by U}
        {
          \For{Pairs of indices of X Matrices in Pauli string in sorted in ascending order}
          {
            Apply a CNOT to this pair with control on the larger index
          }
          Diagonalise remaining matrix with an Rx or H\\
          Append this sequence to U
        }
        
      }
      return U
      
    \caption{Diagonalise commuting Pauli strings}
\end{algorithm}

\begin{figure}
    \centering
    \includegraphics[width=\columnwidth]{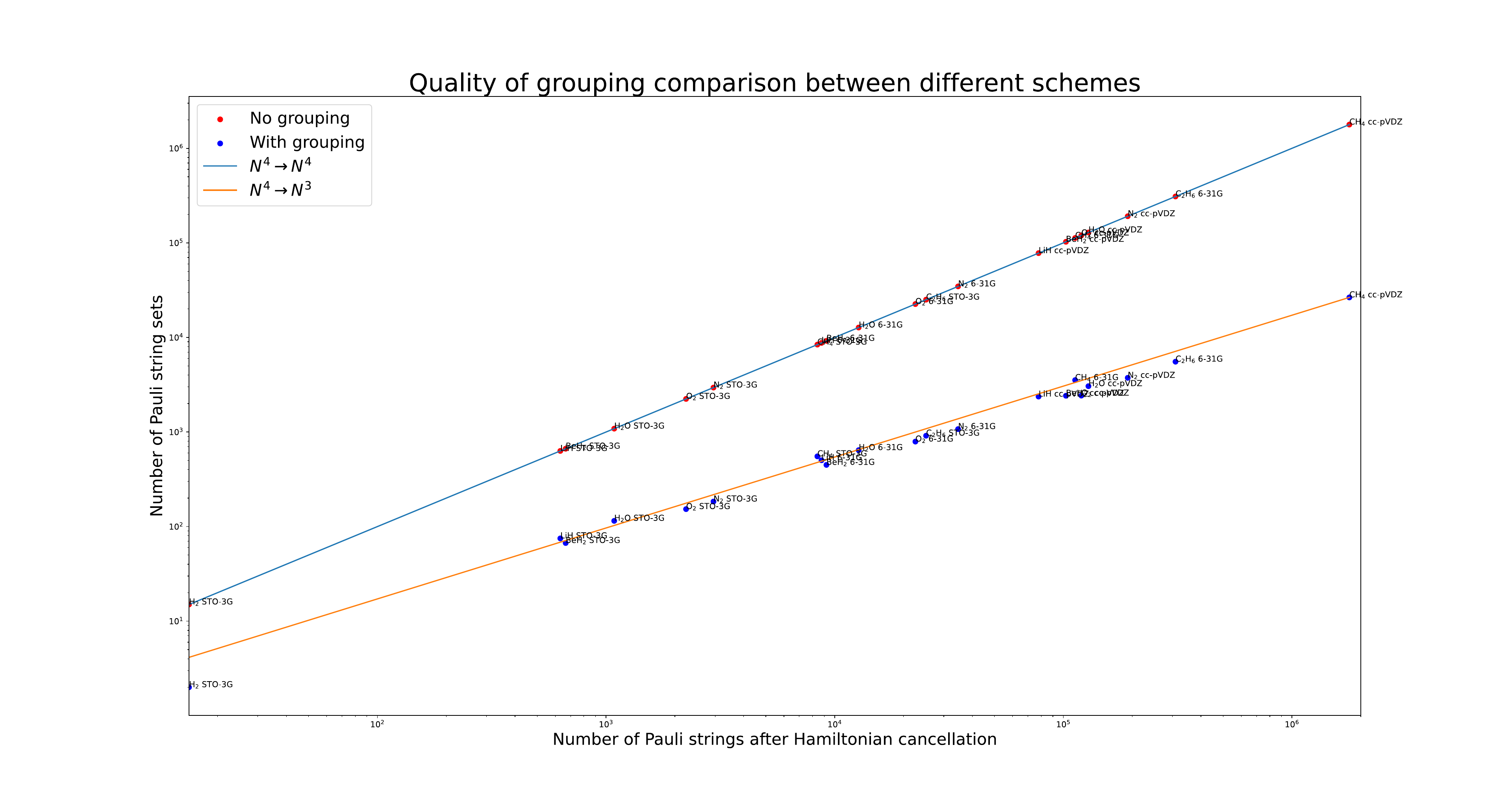}
    \includegraphics[width=\columnwidth]{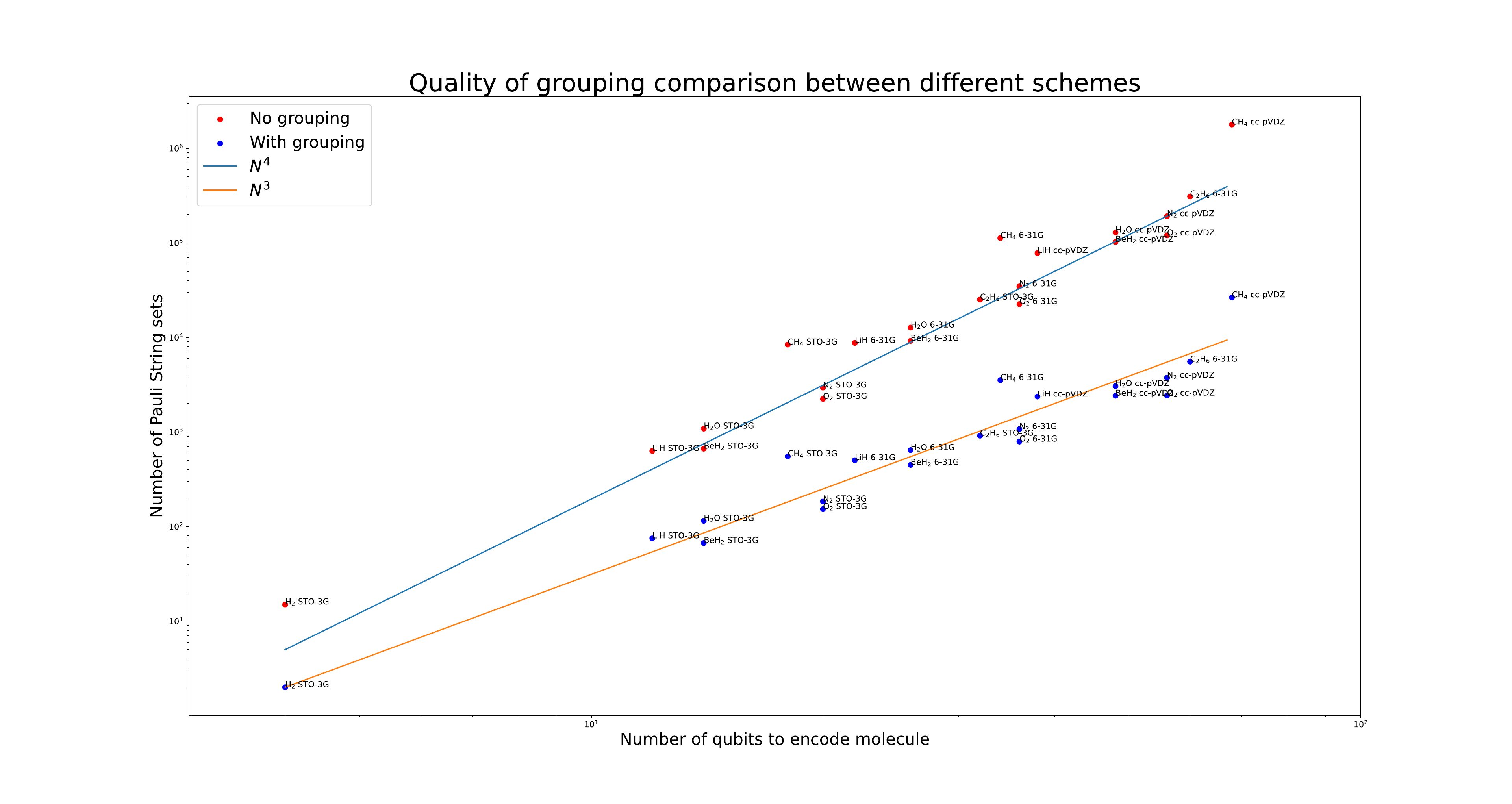}
    \caption{Top, number of commuting Pauli string sets, after cancellation and removal of zero coefficient terms, as a function of total number of Pauli strings. Bottom, number of commuting Pauli string sets as a function of number of qubits. The total number of Pauli strings is $O(N^4)$ the orange curve shows that Baranyai grouping has achieved $O(N^3)$ measurement complexity. It can be seen that this reduction is present in real Hamiltonians, not just idealised ones. The scatter around the trendlines is due to the number of zeros from symmetry being different in different molecules}
    \label{fig:BaranyaiGroupingResult}
\end{figure}

\subsection{Other Mappings}
Linear transformations between bases preserve commutativity of matrices acting on a vector space.\cite{choi_linear_1987}. This can also be seen from Equation \ref{eq:CommutePreserved} where $[A,B]$ denotes the commutator between $A$ and $B$, and $R$ denotes the coefficient map. 
\begin{equation}
\begin{gathered}
    \relax{[RAR^{-1},RBR^{-1}]} = R[A,B]R^{-1}
    \label{eq:CommutePreserved}
\end{gathered}
\end{equation}
Since the JW mapping is related to the Fock Space in a linear, invertible fashion, this grouping scheme is valid for any mapping that is a linear mapping of the Fock space. For example the Bravyi--Kitaev mapping, while a more complicated mapping from Fock space to qubit space, is still one to one and linear in the coefficients\cite{tranterJWBKComparison}.
\section{Results}
This solution was applied to 21 different molecular Hamiltonians for different molecules (H\textsubscript{2} -- C\textsubscript{2}H\textsubscript{6}) and basis sets (STO-3G -- cc-pVDZ)), and the results are shown in Figure \ref{fig:BaranyaiGroupingResult} and tabulated in Table \ref{Supp-tab:BaranyaiGroupingResult}. The grouping showed a reduction in the number of terms from $O(N^4)$ to $O(N^3)$ on real, chemically relevant, Hamiltonians. The partitioning algorithm completed in a short time on a single core in Python, with the longest partitioning taking 2.5 hours and others taking minutes. Sparse linearly-scaling Hamiltonians were also tested and the resultant $O(1)$ group count is shown in Figure \ref{Supp-fig:LinearHamiltonian} in the Supplementary Information. The code is available as part of the \textit{FS-VQE} framework\cite{FS-VQE,Qiskit}. 
\section{Conclusion}
Here we showed that the Baranyai grouping produces small sets that scale with sparsity in the Hamiltonian. The algorithm completes in polynomial time and does not require significant computational effort. The implications of Hamiltonian partitioning into commuting sets can be used for simplifying VQE Pauli gadgets\cite{Peruzzo2014,tazi2023folded}, or Trotter expansions of Hamiltonian evolution circuits\cite{leadbeater2023nonunitary} by allowing multiple Pauli strings to be simultaneously diagonalised, which is necessary to perform the Rz rotations, reducing the number of diagonalising matrices and significantly lowering the depth of the circuit. Each commuting set generates other Pauli Strings which could be measured to reduce the variance\cite{choi_improving_2022} or Trotter error\cite{choi_measurement_2023}.
Anti-commuting sets can also be simultaneously measured, by non-Clifford rotations to a single Pauli string in the set\cite{zhao_measurement_2020}, so by partitioning terms into either fully commuting or fully anti-commuting sets even fewer sets could be created. 

\section{Acknowledgements}
The authors would like to give a special thanks to Lila Cadi Tazi for her scientific contribution, and to the entire Thom Group for their helpful discussions, and Nathan Fitzpatrick for helpful comments on the manuscript. We also thank the Royal Society of Chemistry for an RSC Undergraduate Research Bursary for this project. 
\section{Author Declarations}
The authors have no conflicts to disclose.\\
All authors contributed equally.


\bibliography{references}

\end{document}


\preprint{APS/123-QED}

\title{Supplementary Information: Optimised Baranyai grouping of the second quantised Hamiltonian}

\author{Bence Csakany}
\author{Alex Thom}%

\maketitle
\section{Measurement Scaling for linearly\textbf{} Hamiltonians}
\begin{figure*}
    \centering
	\includegraphics[width=\linewidth]{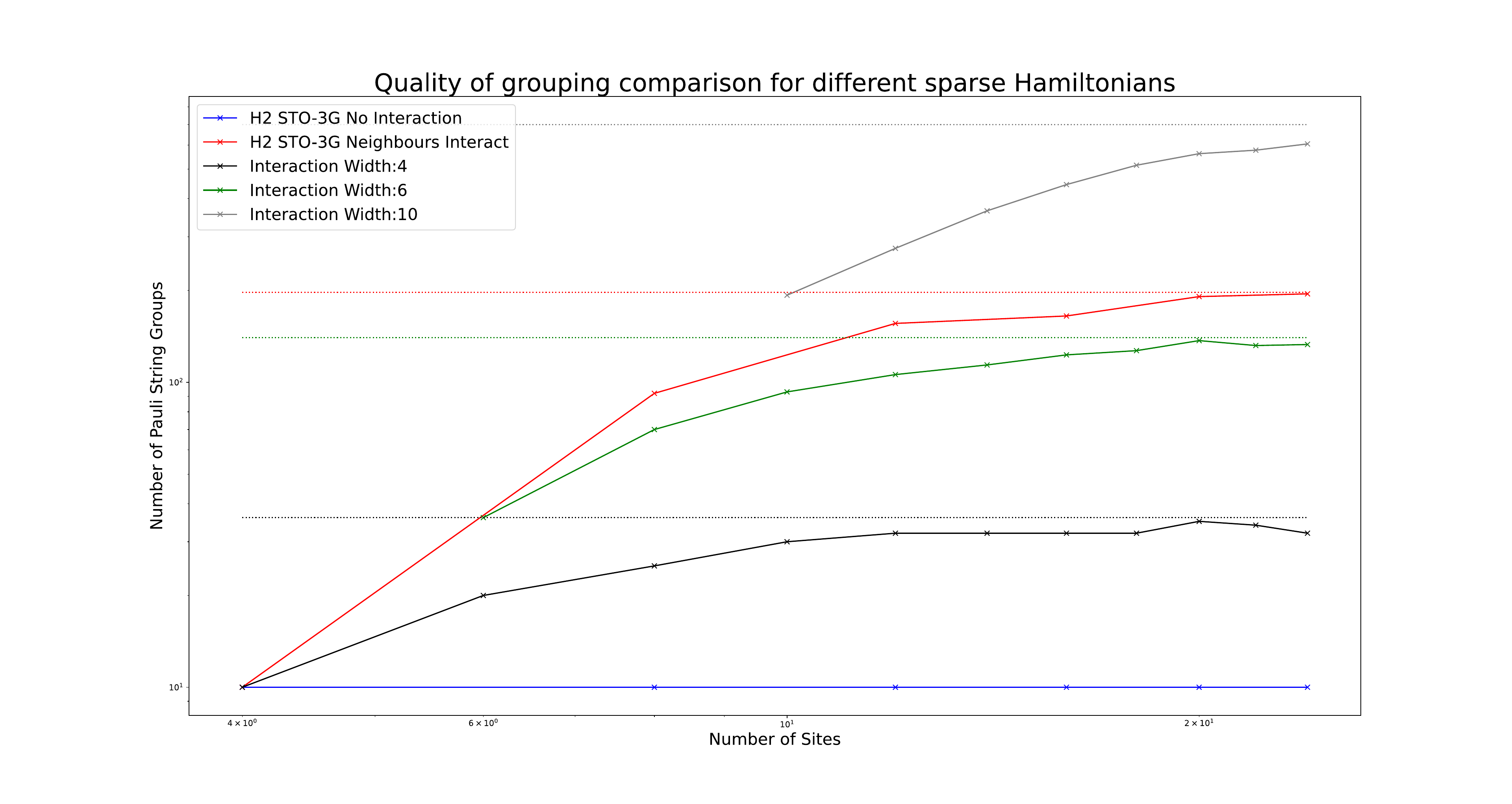}
    \caption{Constant scaling relation for the Hamiltonian of molecules in a line sorted into to commuting sets. Red, blue are a line of H\textsubscript{2} in the STO-3G basis with neighbours interacting/not interacting respectively. Black, green, gray are a string of orbitals in a line with the specified interaction width determining the range of interaction. It can be seen that after a short growth, the number of commuting sets stays constant. The dotted lines are constant and show the asymptotic behaviour. The fluctuation is due to the ordering after Baranyai sorting changing between dimensions}
    \label{fig:LinearHamiltonian}
\end{figure*}

Linearly scaling Hamiltonians can be formed by restricting the range of interaction between neighbouring localized molecular orbitals for a system arranged in a line. A line of 1--8 H\textsubscript{2} molecules with localised orbitals was studied, without cancellation of Pauli strings in the Hamiltonian. Figure \ref{fig:LinearHamiltonian} shows this in action. For example a line of H\textsubscript{2} molecules in the minimal basis would approach the "H\textsubscript{2} STO-3G No Interaction" description as the overlap and interaction between molecules vanishes in the far-separated limit. Figure \ref{fig:LinearHamiltonian} shows that the scaling tends to $O(1)$.  A line of orbitals was also studied in Figure \ref{fig:LinearHamiltonian} and ``Interaction Width'' specifies the range of interaction allowed. For example a width of 4 means orbital 5 interacts with 2,3,4,5,6,7,8.
\section{Reduction in measurement Cost}
Table \ref{tab:BaranyaiGroupingResult} shows the reduction in the number of measurements i.e. number of commuting Pauli string sets, after cancellation and removal of zero coefficient terms. Table \ref{tab:BaranyaiGroupingResult} also shows the number of Pauli strings in the entire Hamiltonian and the number of qubits needed for the molecule--basis combination. The total number of Pauli strings is $O(N^4)$ and the total number of commuting sets is $O(N^3)$. It can be seen that this reduction is present in real Hamiltonians, not just idealised ones.

\section{General Diagonalisation for excitations in Baranyi grouping}
\label{sec:DiagonalisationExample}
\subsection{Example 1}
Consider the disjoint double excitations $a_1^\dag a_2^\dag a_7a_8$ and $a_5^\dag a_6^\dag a_4 a_5$, Taking the `real' part we get a set of 16 Pauli strings, however it is enough to consider only 1 from each excitation (for this example). The choice is arbitrary. After sorting our set becomes 
\newcommand{\blue}[1]{\underline{#1}}
\newcommand{\red}[1]{\overline{#1}}
\[\{II\red{YYXX}II,XX\red{ZZZZ}YY\}\]
where the over(under)-bars are showing the modified elements.  Applying CNOT gate to elements \{3,4\} \{4,5\} \{5,6\} yields
\[\{II\red{ZIIX}II,XX\red{ZIZI}YY\}\]
Applying $H$ to element 6 yields
\[\{IIZII\red{Z}II,\blue{XX}ZIZI\blue{YY}\}\]
The first one is now diagonalised. Applying CNOT to elements \{1,2\}, \{2,7\}, \{7,8\} yields
\[\{IIZIIZII,\blue{II}ZIZI\blue{ZX}\}\]
Applying $H$ to element 8 yields
\[\{IIZIIZII,IIZIZIZ\blue{Z}\}\]
Which is now diagonal. 

\subsection{Example 2}
Diagonalise the set resulting from a repeating index excitation 
\[\{XIX,YZY,IXX,ZYY,XXI,YYZ\}\]
For ease of explanation the set can be reduced to its generators and it suffices to diagonalise these.
\[\{XIX,IXX,ZZZ\}\]
Arranging in sorted order,
\[\{XXI,XIX,ZZZ\}\]
applying a CNOT to element 1,2 yields,
\[\{IXI,XIX,ZIZ\}\]
applying a $H$ to element 2 yields,
\[\{IZI,XIX,ZIZ\}\]
repeating the above for the last non-diagonal Pauli string in the set,
\begin{gather*}
\{IZI,XII,IIZ\}\\
\{IZI,ZII,IIZ\}
\end{gather*}
which is now diagonal. This set did not fit the structure of the other excitations, however because the generators (excluding $ZZZ$) do fit the requirements the algorithm still worked.

\section{Factorisable Double Excitations}
\label{sec:FactorisableDoubleExample}
An example for 8 spin-orbital system where we define indices $\{0,1,2,3\}$ are $\alpha$ spin and $\{4,5,6,7\}$ are $\beta$ spin:
Let $t_{i,j,k,l} = a_i^\dag a_j^\dag a_k a_l + a_k^\dag a_l^\dag a_i a_j$
Therefore the $\alpha$($\beta$) double grouping is:
\[\{t_{0,1,2,3},t_{4,5,6,7}\}\]
A cross double set for a given $\alpha$ index choice is as follows:
\[\{t_{0,1,4,5},t_{0,1,4,6},t_{0,1,4,7},t_{0,1,5,6},t_{0,1,5,7},t_{0,1,6,7}\}\]
this set would be repeated for all choices of $\alpha$ indices. Therefore a packing might be:
\begin{gather*}
    \{t_{0,1,4,5},t_{2,3,6,7}\}\text{ and }\\
    \{t_{0,1,6,7},t_{2,3,4,5}\}
\end{gather*}
where the grouping $\{(0,1),(2,3)\}$ was chosen for the $\alpha$ indices and $\{(4,5),(6,7)\}$ chosen for the $\beta$ indices. The two packings are related by the rotation in the $\beta$ indices. 

\section{Proof that Disjoint Indexes commute}
\label{app:disjointCommute}
Consider the commutator $[a_i^\dag a_j,a_k^\dag a_l]$ where $i,j,k,l$ are unique.
Using the properties of fermionic creation and annihilation operators:
\begin{gather*}
\{a_i,a_j\}=\{a_i^\dag, a_j^\dag\} = 0\\\{a_i,a_j^\dag\} = \delta_{ij}    
\end{gather*}
where $\{A,B\} \equiv AB+BA$\\
We can permute $a_i^\dag a_j a_k^\dag a_l$ into $a_k^\dag a_l a_i^\dag a_j$ using an even number of 2-cycles. Since each two cycle incurs a sign of $-1$ from the above relations, The overall permutation has a sign of $+1$ Therefore
\begin{equation}
    a_i^\dag a_j a_k^\dag a_l = a_k^\dag a_l a_i^\dag a_j
    \label{eq:disjointCommute}
\end{equation}

Since higher order excitations are composed of an even number of creation/annihilation operators, an even number of 2-cycles is also needed so these also commute.
It remains to show that the Pauli Strings comprising an excitation operator all \textbf{individually} commute not just their sums. 
Two Pauli Strings either commute or anticommute i.e.
\[P_1 P_2 = k_{12} P_2 P_1\]
Where $k_{12} = \pm 1$
It will be important that Pauli matrices ($\Pi = \{X,Y,Z,I\}$) form a basis for the 2x2 matrices and the tensor product $\Pi\otimes\Pi$ also forms a basis for the 4x4 matrices. i.e. $\Pi^{\otimes n}$, the n-dimensional tensor product of Pauli matrices, is a basis. It also forms a group.\\

For two Pauli strings from excitations with disjoint indices the set of paulis created by
\[\{P^{(1)}_iP^{(2)}_j\}\]
where $P^{(1)}_i$ denotes the $i$\textsuperscript{th} Pauli string in the $1$\textsuperscript{st} excitation. The following is true
\begin{equation}
    P^{(2)}_jP^{(1)}_i \neq P^{(2)}_kP^{(1)}_l,\ \forall j \neq k\ \text{or}\ i\neq l
    \label{eq:NoClashPaulis}
\end{equation}
Suppose for sake of contradiction the opposite were true.
\begin{equation}
\begin{gathered}
     P^{(2)}_jP^{(1)}_i = P^{(2)}_kP^{(1)}_l\\
    \implies P^{(1)}_i P^{(1)}_l = P^{(2)}_jP^{(2)}_k \\
    \label{eq:NoClashPaulisContra}
\end{gathered}
\end{equation}
For some $i,j,k,l$ where $j \neq k, i\neq l$\\
In JW (Jordan-Wigner) mapping $P^{(1)}_i P^{(1)}_l$ and $P^{(2)}_jP^{(2)}_k$ are both diagonal Pauli strings with $Z$ matrices only on the active indices. All the parity $Z$ matrices have cancelled to give $I$ matrices and each active index either has an $X$ or $Y$ therefore all products lead to $\{Z,I\}$. Since the active indices are disjoint, $P^{(1)}_i P^{(1)}_l$ and $P^{(2)}_jP^{(2)}_k$ have $Z$ matrices on different indices.\\
Therefore cannot be equal, unless they are both $I^{\otimes n}$ (No $Z$ Matrices).\\
\begin{gather*}
    \implies P^{(1)}_i P^{(1)}_l = P^{(2)}_jP^{(2)}_k = I^{\otimes n}\\
    \iff P^{(1)}_i = P^{(1)}_l\ and\ P^{(2)}_j = P^{(2)}_k\\
    \iff i = l\ \text{and}\ j = k, \text{Contradiction}
\end{gather*}
Where the last line follows from $P^{(1)}_i$ being self-inverse and $P^{(1)}_i \neq P^{(1)}_l,\ \forall i \neq l$\\
Therefore, equation \ref{eq:NoClashPaulisContra} must be false, therefore \ref{eq:NoClashPaulis} is true. \\

In the above JW mapping was assumed however if Equation \ref{eq:NoClashPaulis} holds for the mapping of consideration, then the proof below follows. If not then there is a linear combination of Pauli strings which form commuting sets given by:
\[RP^{\text{JW}}R^{-1}\]
where $P^{\text{JW}}$ represents the single Pauli string in JW that maps to the linear combination in the arbitrary mapping. The diagonalising matrix is given by $RU$. This follows from:
\begin{gather*}
    R^{-1} P^\text{mapped}R = P^\text{JW}\\
    U^{-1}R^{-1} P^\text{mapped}RU = U^{-1}P^\text{JW}U = P^{\text{diag}}\\
\end{gather*}
Therefore as described in the main text the set can still be simultaneously diagonalised.\\

Considering the product of two excitation operators and using equation \ref{eq:disjointCommute}
\begin{gather*}
\sum_{ij} P^{(1)}_iP^{(2)}_j = \sum_{ij} k_{ij}P^{(2)}_jP^{(1)}_i = \sum_{ij}P^{(2)}_jP^{(1)}_i\\
\implies \sum_{ij} (k_{ij}-1) P^{(2)}_jP^{(1)}_i = 0
\end{gather*}
since 
\begin{equation}
    P_1 P_2 \neq 0\ \forall P_1,P_2 \in \Pi^{\otimes n}
    \label{eq:NoZeroProductPaulis}
\end{equation} 
and using equation \ref{eq:NoClashPaulis}. Since $\Pi^{\otimes n}$ is a basis
\begin{equation}
    P^{(2)}_k P^{(1)}_l \neq \sum_{i\neq k,j\neq l} c_{ij} P^{(2)}_jP^{(1)}_i,\iff c_{ij} = 0,\ \forall i,j
    \label{eq:PauliLinearlyIndependent}
\end{equation}
Therefore each term must be individually 0, so the only solution is $k_{ij} = 1$. Therefore each individual Pauli string commutes $\square$.\\

To finish we provide a counterexample for equation \ref{eq:NoClashPaulis}. Consider $A = X\otimes Y + I\otimes Y$ and $B = X \otimes X + I \otimes X$
we see that $[A,B] = 0$, infact $AB = 0$
but none of the terms across the sets commute.\\

Infact, there is a basis in which $H$ the Hamiltonian is diagonal. $RHR^{-1} = \Lambda = \sum_i c_i P_i$ where $P_i$ are diagonal Pauli strings and therefore all commute. So the basis we are viewing the problem in (the Fock space) is already a basis change from simultaneously diagonalisable Pauli matrices, $P_i$ are trivially diagonalisable if they are already diagonalised. Therefore, it is not surprising that a different basis change could lead to Pauli strings representing a matrix in the Fock space, which do commute, being transformed into a non mutually commuting set.

\section{Baranyai grouping where k does not divide n}
\label{app:LargerSetsProof}
For the case where we need to pack $\binom{n}{k}$ elements but $r = n\ \text{mod}\ k \neq 0$, there is a choice whether to embed it into a larger $N = n+k-r$ or smaller $N = n-r$ space.
To encompass all $\binom{n}{k}$ elements you need $\binom{n}{r}$ smaller $N$ sets therefore in total there are:
\[\binom{n-r-1}{k-1} \binom{n}{r}\]
sets. If embedding into a larger $N$ then there are:
\[\binom{n+k-r-1}{k-1}\]
number of sets. 
The fraction of the smaller $N$ sets to larger $N$ sets is:
\begin{gather}
    R = \frac{\binom{n-r-1}{k-1} \binom{n}{r}}{\binom{n+k-r-1}{k-1}}\\
    = \frac{(n-r-1)!\ n!}{(n-r-k)!\ r!\ (n+k-r-1)!}\\
    = \frac{n!}{(n-r-k)!}/\frac{(n+k-r-1)!\ r!}{(n-r-1)!}\\
    \approx \frac{n^{r+k}}{n^k\ r!} > 1\ \text{for}\ n >> r
\end{gather}
Therefore embedding into larger $N$ yields fewer sets.

\begin{table*}
    \centering
    \begin{tabular}{|c|c|c|c|c|}
    \hline
        Molecule & Basis & Num. of Commuting Groups & Num. of Pauli strings & Num. of Qubits \\
    \hline
    H$_2$&STO-3G&2&15&4\\
    LiH&STO-3G&75&631&12\\
    BeH$_2$&STO-3G&67&666&14\\
    H$_2$O&STO-3G&115&1086&14\\
    CH$_4$&STO-3G&553&8400&18\\
    N$_2$&STO-3G&184&2951&20\\
    O$_2$&STO-3G&153&2239&20\\
    C$_2$H$_6$&STO-3G&914&25109&32\\
    &&&&\\
    LiH&6-31G&503&8758&22\\
    BeH$_2$&6-31G&449&9204&26\\
    H$_2$O&6-31G&643&12732&26\\
    CH$_4$&6-31G&3538&112632&34\\
    N$_2$&6-31G&1074&34655&36\\
    O$_2$&6-31G&792&22543&36\\
    C$_2$H$_6$&6-31G&5542&309499&60\\   
    &&&&\\
    LiH&cc-pVDZ&2370&77930&38\\
    BeH$_2$&cc-pVDZ&2419&102681&48\\
    H$_2$O&cc-pVDZ&3054&128785&48\\
    N$_2$&cc-pVDZ&3743&191321&56\\
    O$_2$&cc-pVDZ&2427&119909&56\\
    CH$_4$&cc-pVDZ&26504&1782039&68\\
    \hline
    \end{tabular}
    \caption{Table showing the number of commuting sets achieved for a given molecule, qubit count and basis. The table also shows the number of Pauli strings present in the entire Hamiltonian. The reduction in number of measurements can be seen as the number of Pauli Strings increases faster than the number of commuting groups in all molecules and bases.}
    \label{tab:BaranyaiGroupingResult}
\end{table*}